\documentclass[aps,pra,twocolumn,groupedaddress]{revtex4-1}
\usepackage{graphicx}
\usepackage{wrapfig}
\usepackage{latexsym}
\usepackage{amssymb}
\usepackage{amsfonts,color}
\usepackage{float}
\usepackage{amsmath}

\newcommand{\be}{\begin{equation}}
\newcommand{\ee}{\end{equation}}
\newcommand{\ba}{\begin{align}}
\newcommand{\ea}{\end{align}}

\begin{document}


\title{Contact interactions and Kronig-Penney Models in Hermitian and ${\cal PT}$ symmetric Quantum Mechanics}
\author{Foster Thompson$^{(1)}$, Katherine Brown$^{(2)}$, Harsh Mathur$^{(1)}$ and Kristin McKee$^{(1)}$}
\vspace{2mm}
\affiliation{$^{(1)}$Department of Physics, Case Western Reserve University, Cleveland, Ohio 44106}

\vspace{2mm}
\affiliation{$^{(2)}$Department of Physics, Hamilton College, Clinton, NY 13323}



\date{\today}

\begin{abstract}

The delta function potential is a simple model of zero-range contact interaction in non-relativistic quantum mechanics in 
one dimension. The Kr\"{o}nig-Penney model is a one-dimensional periodic array of delta functions and provides a simple 
illustration of energy bands in a crystal. Here we investigate contact interactions that generalize the delta function potential
and corresponding generalizations of the Kr\"{o}nig-Penney model within conventional and ${\cal PT}$ symmetric quantum mechanics. In 
conventional Hermitian quantum mechanics we determine the most general contact interaction compatible with 
self-adjointness and in ${\cal PT}$ quantum mechanics we consider interactions that respect symmetry under the transformation ${\cal PT}$ 
where ${\cal P}$ denotes parity and ${\cal T}$ denotes time reversal. In both cases we find that the most general interaction has four 
independent real parameters and depending on the values of those parameters the contact interaction can support zero, one 
or two bound states. By contrast the conventional delta function can only support zero or one bound state. In the ${\cal PT}$ 
symmetric case moreover the two bound state energies can be both real or a complex conjugate pair. The transition from real 
to complex bound state energies corresponds to the spontaneous breaking of ${\cal PT}$ symmetry. The scattering states for the ${\cal PT}$ 
symmetric case are also found to exhibit spontaneous breaking of ${\cal PT}$ symmetry wherein the eigenvalues of the non-unitary S-
matrix depart the unit circle in the complex plane. We also investigate the energy bands when the generalized contact 
interactions are repeated periodically in space in one dimension. In the hermitian case we find that the two bound states result 
in two narrow bands generically separated by a gap. These bands intersect at a single point in the Brillouin zone as the 
interaction parameters are varied. Near the intersection the bands form a massless Dirac cone. In the ${\cal PT}$ symmetric case we 
find that as the parameters of the contact interaction are varied the two bound state bands undergo a ${\cal PT}$ symmetry breaking 
transition wherein the two band energies go from being real to being a complex conjugate pair. The ${\cal PT}$ symmetric Kr\"{o}nig-
Penney model provides a simple soluble example of the transition which has the same form as in other models of ${\cal PT}$ symmetric 
crystals.

\end{abstract}

\pacs{}

\maketitle

\section{Introduction \label{introduction}}

A fundamental principle of quantum mechanics is that operators which correspond to
observable quantities, most notably, the Hamiltonian, must be Hermitian \cite{hermiticity}. 
Recently there has been
a surge of interest in operators that are non-hermitian but respect the combined 
symmetry ${\cal PT}$ where ${\cal P}$ denotes parity and ${\cal T}$ is time reversal \cite{benderprl, benderrev}. 
In classical optics it has proved possible to fabricate materials with alternating regions
of gain and loss that demonstrate many novel optical properties (for recent reviews see \cite{photo1, photo2}). 
In such systems the equation that governs the propagation of 
electromagnetic waves can be engineered to have the form of a Schr\"{o}dinger
equation with ${\cal PT}$ symmetry. In order to build intuition for wave propagation
in these materials it is therefore relevant to consider simple models of ${\cal PT}$ 
quantum mechanics. In this paper we construct the ${\cal PT}$-symmetric generalizations 
of two models well-known from conventional Hermitian quantum mechanics: the delta function
potential and the simplest model of a periodic crystal, the Kr\"{o}nig-Penney model. 


The delta function is a widely used model of a zero-range contact interaction in 
quantum mechanics. Rigorously it is a viable model of contact interaction only in 
one dimension. 
In higher dimensions the ideal
delta function potential is invisible and it is better to treat contact interactions as 
modified boundary conditions \cite{jackiw}. Here we show that even in one
dimension it is helpful to model a contact interaction as a boundary condition;
adopting this point of view we find that even in hermitian quantum mechanics
in one dimension the delta function is merely a special case of the most general
allowed contact interaction.
Quite different forms of contact interaction emerge when we relax the conditions of
hermitian self-adjointness but instead impose the requirement of ${\cal PT}$ symmetry. 
We find that for both the hermitian and ${\cal PT}$ symmetric generalized contact
interactions there can be zero, one or two bound states depending on the parameters
that characterize the 
interaction. In contrast the conventional delta function
can only have zero or one bound states depending on whether the the potential
is attractive or repulsive. 
In the ${\cal PT}$ symmetric case 
when there are two bound states the eigenvalues can be either both
real or a complex conjugate pair depending on the  parameters of the model.
As the parameters pass through a critical value, the real eigenvalues degenerate and enter
the complex plane, behavior that is called the ${\cal PT}$ transition \cite{benderrev}. 
The ${\cal PT}$ transition is accompanied by spontaneous breaking of ${\cal PT}$ symmetry:
although the interaction remains ${\cal PT}$ symmetric, the eigenstates are no longer 
invariant under ${\cal PT}$. 
We also identify
a ${\cal PT}$ transition in the scattering states. In this case the energy is necessarily
real; the transition occurs when the eigenvalues of the non-unitary $S$-matrix 
cease to be unimodular and depart from the unit circle in the complex plane \cite{doug}. 


We generalize the conventional Kr\"{o}nig-Penney model by considering 
a periodic array of generalized contact interactions in one dimension.
In the hermitian case 
the two bound state bands have a
simple cosine dispersion when they are well separated. However when the parameters
of the contact interaction are tuned suitably the bands intersect at an isolated point 
in the Brillouin zone. Near the intersection the band structure is a massless 
Dirac cone. 
This behavior is reminiscent of topological insulators where gap closure
is a phase boundary that separates an ordinary insulator from a topological insulator \cite{kane}. 
Whether that is the case here is a question we leave open for future work. 
For the ${\cal PT}$ symmetric case the two bound state bands 
undergo a ${\cal PT}$ symmetry breaking transition as the parameters are
varied (see fig \ref{fig:bandbreak}). Before the onset of the transition the two bands are entirely real.
After the transition is complete the bands are a complex conjugate pair. 
For intermediate values of the parameters the bands are real over part
of the Brillouin zone and a conjugate pair over the remainder. The 
${\cal PT}$ symmetric Kr\"{o}nig-Penney model thus constitutes a particularly
simple and soluble model that exhibits these generic features of ${\cal PT}$ 
symmetric crystals. The generalized hermitian Kr\"{o}nig-Penney model
may be useful as a description of semiconductor superlattices 
\cite{davies} 
and the ${\cal PT}$ symmetric generalization may be relevant to 
experiments in ${\cal PT}$ optics \cite{photo1, photo2}. 

Boundary conditions that respect ${\cal PT}$ symmetry were first introduced
by Krejcirik {\em et al.} in context of a particle in a box and generalizations thereof 
in a series of papers \cite{krejcirik1, krejcirik2, krejcirik3, krejcirik4, krejcirik5}; see also \cite{dasarthy}. 
The study of periodic ${\cal PT}$ symmetric potentials was initiated
by ref \cite{benderperiod, jones, ahmed}. 
Subsequently 
ref \cite{creol} spurred experimental activity in the field by identifying practical realizations 
in optics and
by discovering novel wave propagation effects in crystals with ${\cal PT}$ symmetry. 
The work of refs \cite{jones, ahmed} is particularly closely related to the present work.
These authors introduced and analyzed a version of the Kr\"{o}nig-Penney model wherein
the periodic potential is piecewise constant. Here by contrast we consider a different Kr\"{o}nig-Penney
model that consists of repetitions of zero range contact interactions that cannot be obtained 
from the models of refs \cite{jones, ahmed} by any limiting procedure. 
Motivated by very different considerations of topology change, quantum gravity and 
many-worlds quantum mechanics the authors of ref \cite{wilczek} have also 
considered hermitian generalizations of the contact interaction. We discuss
the relationship of our results to ref \cite{wilczek} in section \ref{sec:hermitdelta}.

\section{Contact Interaction}

\label{sec:contact}

Consider the textbook problem of a non-relativistic particle of mass $m$ in one 
dimension interacting with a delta function potential $\lambda \delta (x)$ located
at the origin. Rather than treating the delta function as a potential we may 
regard it as a boundary condition that the wave function must satisfy, 
namely, continuity at the origin, 
$\psi ( 0 ^+ ) = \psi ( 0^- )$, 
and discontinuity in the derivative given by 
\begin{equation}
\psi' ( 0^+ ) = \frac{2 m \lambda}{\hbar^2} \psi ( 0^- ) + 
\psi' ( 0^- ).
\label{eq:ordinarydeltabc}
\end{equation}
Viewing the delta function as a boundary condition suggests a more
general model of a contact interaction wherein the wavefunction satisfies the boundary
condition
\begin{eqnarray}
\psi ( 0^+ ) & = & a \psi ( 0^- ) + b \psi' ( 0^- ) \nonumber \\
\psi' ( 0^+ ) & = & c \psi ( 0^- ) + d \psi' ( 0^- )
\label{eq:gendelta}
\end{eqnarray}
where $a, b, c$ and $d$ are complex constants. This is the most general boundary condition
compatible with linearity and the order of the Schr\"{o}dinger equation. 
The conventional delta function is the special case $a = 1, b = 0, c = 2 m \lambda/\hbar^2$ and 
$d = 1$. Below we show
that imposing the requirements of self-adjointness or ${\cal PT}$ 
symmetry powerfully constrain the form of the boundary condition (\ref{eq:gendelta}). However in both
cases boundary conditions more general than the conventional delta function are permissible
and represent new kinds of zero range contact interaction; this is a key finding of the present work. 

In the remainder of this paper we will work in units wherein $\hbar =1$ and the mass of the
particle $m = 1$. 


\subsection{Hermitian quantum mechanics}

\label{sec:hermitdelta}

\subsubsection{The Model}

\label{sec:hermitmodel}

Consider a non-relativistic particle 
in one dimension, free except for a zero range contact potential at the origin.
The inner product of two states $\phi$ and $\psi$ is given by
\begin{equation}
( \phi, \psi ) = \int_{-\infty}^{0^-} d x \; \phi^\ast (x) \psi (x) + \int_{0^+}^{\infty} d x \; \phi^\ast (x) \psi (x).
\label{eq:inner}
\end{equation}
Straightforward integration by parts reveals that the free particle Hamiltonian satisfies
\begin{equation}
(\phi, H \psi) = ( H \phi, \psi ) + {\rm surface} \hspace{2mm} {\rm terms};
\label{eq:formalself}
\end{equation}
hence $H$ is formally self adjoint with respect to the inner product
(\ref{eq:inner}). The surface term at the origin is proportional to
\begin{equation}
\left[ \phi^\ast \psi' - \phi^{\ast '}  \psi \right]_{0^+} - 
\left[ \phi^\ast \psi' - \phi^{\ast '}  \psi \right]_{0^-}.
\label{eq:surface}
\end{equation}
To determine what boundary conditions are compatible with the self adjointness of $H$ we proceed
as follows \cite{goldbart}. We impose the boundary condition given in eq (\ref{eq:gendelta}) 
on $\psi$ and ask what boundary
condition must be imposed on $\phi$ in order to make the surface term vanish. Let us write the boundary
condition on $\phi$ as
\begin{eqnarray}
\phi ( 0^+ ) & = & A \phi ( 0^- ) + B \phi' ( 0^- ) \nonumber \\
\phi' ( 0^+ ) & = & C \phi ( 0^- ) + D \phi' ( 0^- ).
\label{eq:bbar}
\end{eqnarray}
It is then easy to verify that the surface terms in eq (\ref{eq:surface}) will vanish provided
\begin{eqnarray}
A^\ast & = & a/ (ad - bc), 
\nonumber \\
B^\ast & = & b/ (ad - bc), 
\nonumber \\
C^\ast & =  & c/ (ad - bc), 
\nonumber \\
D^\ast & = & d/ (ad - bc).
\label{eq:vanish}
\end{eqnarray}
The operator $H$ is self adjoint when the boundary condition imposed on $\psi$ inexorably 
requires the same boundary condition be imposed on $\phi$ \cite{goldbart}. Hence the boundary conditions
compatible with the self adjointness of $H$ are that
\begin{eqnarray}
a & = & \alpha e^{i \theta} \nonumber \\
b & = & \beta e^{i \theta} \nonumber \\
c & = & \gamma e^{i \theta} \nonumber \\
d & = & \delta e^{ i \theta}.
\label{eq:abcd}
\end{eqnarray}
Here $\alpha, \beta, \gamma$ and $\delta$ are real and satisfy $\alpha \delta - \beta \gamma = 1$ \cite{footnote:gauge}.


In summary the most general form of contact interaction compatible with self-adjointness
is given by eq (\ref{eq:gendelta}) 
with the additional constraint that the matrix of coefficients
\begin{equation}
\left( \begin{array}{cc} 
a & b \\
c & d
\end{array}
\right)
\label{eq:sl2r}
\end{equation}
is an $SL (2, R)$ matrix (i.e. it has real entries and unit determinant) multiplied by a phase. 
The general contact interaction described above is time reversal symmetric for $\theta = 0$ or $\pi$. 
This is because if a wavefunction 
$\psi$ satisfies the boundary condition (\ref{eq:gendelta}) with real coefficients,
then so does its time reversed counterpart
$\psi^\ast$. Parity is respected
only if we impose $a = d$. In that case one
can verify that if $\psi(x)$ satisfies the boundary condition (\ref{eq:gendelta}) then so does ${\cal P} \psi (x) = 
\psi( - x)$. 


To conclude this subsection we discuss the connection of these results to the findings of ref \cite{wilczek}. We can
rewrite eq (\ref{eq:gendelta}) as 
\begin{eqnarray}
\psi (0^+) & = & \frac{ \alpha }{ \gamma } \psi' ( 0^+ ) - \frac{ e^{i \theta} }{ \gamma } \psi' (0^- )
\nonumber \\
\psi (0^-) & = & \frac{e^{- i \theta} }{ \gamma } \psi' ( 0^+ ) - \frac{\delta}{\gamma} \psi' ( 0^- ).
\label{eq:disconnect}
\end{eqnarray}
From eq (\ref{eq:gendelta}) we see that $\alpha = \delta = 1$ and $\beta = \theta = 0$
and $\gamma = 0$ corresponds to zero 
interaction. In this case the wave function
and its derivative are continuous and the wave function is smooth across the origin. 
From eq (\ref{eq:disconnect}) we see that for $\gamma \rightarrow \infty$ (with
$\alpha, \delta$ finite) the positive and negative half lines become disconnected
with Dirichlet boundary conditions applied at the origin on either side. It follows that 
if we fix $\alpha = \delta = 1$ and $\beta = \theta = 0$ then as $\gamma$ goes
from zero to $\infty$ we interpolate continuously from the smooth case to the
disconnected case. This interpolation is the topological transition discussed by
ref \cite{wilczek}. 
Another continuous trajectory through the space of hermitian
boundary conditions is to choose $\alpha = \delta = \cosh s$ and $\beta = \xi \sinh s$
and $\gamma = \sinh s / \xi$ where $\xi$ is a fixed constant and $s$ varies
from $s = 0$ to $s = \infty$. This trajectory starts from zero contact interaction and 
terminates in the disconnection of the two half lines but with the boundary conditions
$\psi (0^+) = \xi \psi ' (0+)$ and $\psi (0^-) = - \xi \psi' (0^-)$ on either side
of the origin in place of Dirichlet boundary conditions. 
These boundary conditions allow for the possibility
of bound states that are confined close to the origin on both sides if $\xi < 0$. 

\subsubsection{Bound states}

We seek a solution of the form 
\begin{eqnarray}
\psi & = & A \exp ( - \kappa x ) \hspace{3mm} {\rm for} \hspace{3mm} x > 0,
\nonumber \\
& = & B \exp ( \kappa x ) \hspace{3mm} {\rm for} \hspace{3mm} x < 0.
\label{eq:boundansatz}
\end{eqnarray}
This solution satisfies the free particle Schr\"{o}dinger equation and has an energy $ - \kappa^2 /2$. 

Application of the boundary condition (\ref{eq:gendelta}) reveals that $\kappa$ must satisfy
\begin{equation}
\beta \kappa^2 + (\alpha + \delta) \kappa + \gamma = 0.
\label{eq:quadratic}
\end{equation}
For the case $\beta \neq 0$ this equation has two roots which can be written in the form
\begin{equation}
\kappa_{\pm} = - \frac{ \alpha + \delta }{2 \beta} \pm \frac{ \sqrt{ 4 + (\alpha - \delta)^2 } }{2 \beta}. 
\label{eq:roots}
\end{equation}
(Here we have made use of the condition $\alpha \delta - \beta \gamma = 1$.) 
Thus both roots are necessarily real. 
For the root to correspond to a viable bound state it must also be positive. Depending 
on the choice of $\alpha, \beta$ and $\delta$ it is possible that 
zero, one or both of the roots are positive. Thus in contrast to the
conventional delta function which can only have zero or one bound states, our generalized zero range
potential is capable of having two bound states. 

The case $\beta = 0$ includes the conventional delta function as a special case. 
For this case eq (\ref{eq:quadratic}) is linear and has just one root
\begin{equation}
\kappa = - \frac{\alpha \gamma}{\alpha^2 + 1}.
\label{eq:root}
\end{equation}
Here we have made use of $\alpha \delta = 1$ to write the root in a particularly transparent form. 
Evidently the root corresponds to a bound state if $\alpha \gamma < 0$. 

Note that the bound states are independent of the phase $\theta$. This is because the bound states
decay exponentially as $x \rightarrow \pm \infty$; hence in this case it is permissible to gauge away 
the phase $\theta$ by a large gauge transformation. 

Finally we note for later use that eq (\ref{eq:quadratic}) suggests an alternative way to parametrize
a hermitian contact interaction using $\beta$ and the 
two real roots of the quadratic form eq (\ref{eq:quadratic}) as independent parameters. 
Denoting the roots $\kappa_1$ and $\kappa_2$ with $\kappa_1 > \kappa_2$ we can easily 
reconstruct $\alpha + \delta = - \beta (\kappa_1 + \kappa_2)$ and $\gamma = \beta \kappa_1 \kappa_2$
from the fact that $\kappa_1$ and $\kappa_2$ are roots of (\ref{eq:quadratic}). In order to reconstruct
$\alpha - \delta$ we make use of $\alpha \delta - \beta \gamma = 1$ to show that 
$\alpha - \delta = \pm \sqrt{ \beta^2 ( \kappa_1 - \kappa_2 )^2 - 4 }$. Hence we see 
that we may use $\beta, \kappa_1$ and $\kappa_2$ as an alternative set of parameters provided
we also specify the sign of $\alpha - \delta$ and respect the constraint that $ ( \kappa_1 - \kappa_2 ) | \beta | \geq 2$. 

\subsubsection{Scattering states}

Next we turn to positive energy scattering states. A state that is incoming from the left has the behavior
\begin{eqnarray}
\psi & = & e^{i k x} + r e^{- i k x} \hspace{3mm} {\rm for} \hspace{3mm} x < 0,
\nonumber \\
& = & t e^{i k x} \hspace{3mm} {\rm for} \hspace{3mm} x > 0. 
\label{eq:scatt}
\end{eqnarray}
The scattering coefficients $t$ and $r$ are determined by imposing the boundary condition
eq (\ref{eq:gendelta}). For simplicity let us suppose initially that the phase $\theta$ is zero. 
For the case $\beta \neq 0$ the transmission coefficient
\begin{equation}
t = \frac{2 i k}{ \beta (k - i \kappa_+) (k - i \kappa_-) }
\label{eq:tleftb}
\end{equation}
and for $\beta = 0$
\begin{equation}
t = \frac{1}{\alpha+\delta} \frac{2 k}{ k - i \kappa}
\label{eq:tleftnob}
\end{equation}
Thus in each case the transmission coefficient has poles along the positive imaginary axis
in the $k$ plane at locations determined by the bound states, consistent with the general analytic properties of the
$S$-matrix in quantum mechanics.
One can similarly analyze a scattering state that is incoming from the right. The scattering coefficients in this
case are denoted $t'$ and $r'$. 
For the sake of brevity we omit expressions for $r$ and $r'$ but note that explicit calculation confirms that 
\begin{equation}
S = 
\left(
\begin{array}{cc}
t & r' \\
r & t 
\end{array} 
\right) 
\label{eq:smatrix}
\end{equation}
is unitary as expected on general grounds. Furthermore $t = t'$ which is a general consequence of time 
reversal symmetry combined with the unitarity of the $S$-matrix. On the other hand $r \neq r'$ unless
$\alpha = \delta$ ensuring that parity is also a symmetry. 

Now let us consider the effect of the phase angle $\theta$ which has so far been set equal to zero in this
subsection. By explicit calculation or use of a gauge argument one can show that the $S$ matrix now becomes
\begin{equation}
S = 
\left(
\begin{array}{cc}
t e^{- i \theta} & r' \\
r & t e^{i \theta}
\end{array} 
\right) 
\label{eq:sphase}
\end{equation}
Hence as claimed the $S$ matrix has a non-trivial dependence on the phase $\theta$. Moreover the transmission
coefficient for incidence from the left and right is no longer the same once time reversal symmetry is broken. 

\subsection{${\cal PT}$ quantum mechanics}

\label{sec:ptdelta}

\subsubsection{The model}

\label{sec:model}

In ${\cal PT}$ quantum mechanics we eschew the condition of self-adjointness with respect to the 
inner product expressed in eq (\ref{eq:inner}) 
but instead require 
the Hamiltonian to respect ${\cal PT}$ symmetry. In the present context we require that if $\psi$ satisfies
the boundary condition eq (\ref{eq:gendelta}) then so should ${\cal PT} \psi(x) = \psi^\ast (-x)$. Straightforward
analysis shows that this condition is met provided the coefficients 
are given by eq (\ref{eq:abcd}) together with the conditions
(i) $\beta$ and $\gamma$ are real (ii) $\alpha = \delta^\ast$ and (iii) $\alpha \delta - \beta \gamma = 1$. 
Thus the primary departure from the
Hermitian case is that $\alpha$ and $\delta$ are no longer required to be real but are required to be a complex conjugate
pair. Hence the number of independent parameters that specify the interaction is the same in both cases. 

If we impose the condition that both ${\cal P}$ and ${\cal T}$ symmetry should be separately respected
we find exactly the same conditions as in the Hermitian case with ${\cal P}$ and ${\cal T}$ symmetry; 
namely, that $a, b, c, d$ must be real, $a= d$ and $ad - bc = 1$. For the record we note that if we impose only
${\cal T}$ symmetry we obtain the condition that $a, b, c, d$ are real. If we impose only ${\cal P}$ 
symmetry we need $a = d$ and $ad - bc = 1$ but there is no restriction to real values for any of the
coefficients. 

\vspace{3mm}

\subsubsection{Bound states}

We now turn to the analysis of the bound states of a ${\cal PT}$ symmetric
contact interaction. The analysis closely parallels that for the hermitian case. 
We seek states of the same form as in the hermitian case 
given by eq (\ref{eq:boundansatz}) 
but this time we no longer require $\kappa$ to be real. We do need the real part of
$\kappa$ to be positive to ensure that the solution vanishes for $x \rightarrow \pm \infty$. 
The ansatz  (\ref{eq:boundansatz}) remains a solution to the
Schr\"{o}dinger equation with energy $- \kappa^2/2$. Application of the boundary condition
eq (\ref{eq:gendelta}) again reveals that $\kappa$ satisfies eq (\ref{eq:quadratic}) but the subsequent
analysis departs from the hermitian case. 

First let us consider the case $\beta \neq 0$. 
Taking into account that $\delta = \alpha^\ast$ we may write the
roots of eq (\ref{eq:quadratic}) in the transparent form
\begin{equation}
\kappa_\pm = - \frac{\alpha_R}{\beta} \pm \frac{1}{\beta} \sqrt{ 1 - \alpha_I^2 }
\label{eq:kappafloor}
\end{equation}
where $\alpha_R$ and $\alpha_I$ are respectively the real and imaginary parts of $\alpha$. 
We now separately consider the cases $| \alpha_I | > 1$ and $| \alpha_I | \leq 1$. 
(i) For $| \alpha_I | > 1$ the roots
$\kappa_\pm$ are a complex conjugate pair. For $- \alpha_R/ \beta > 0$ the real parts of both $\kappa_+$
and $\kappa_-$ are positive and hence there are two bound states. The energies of the two bound
states are complex conjugates of each other. On the other hand for $- \alpha_R/\beta < 0$ the real parts of
$\kappa_\pm$ are negative and hence there are no bound states.
(ii) For $| \alpha_I | < 1$ the roots $\kappa_\pm$ are both real. 
The roots correspond to bound states
only if their real parts are positive. 
It follows from eq (\ref{eq:kappafloor}) that there can be zero,
one or two bound states depending on the values of $\alpha_R, \alpha_I$ and $\beta$. 

Next consider the case that $\beta = 0$. In this case eq (\ref{eq:quadratic}) is linear and there is
only one root
\begin{equation}
\kappa = - \frac{\gamma}{2 \alpha_R}.
\label{eq:oneroot}
\end{equation}
For $- \gamma/ 2 \alpha_R > 0$ this root corresponds to a bound state; otherwise there are zero bound states. 

\begin{figure}[h]
\includegraphics[width=3.25in]{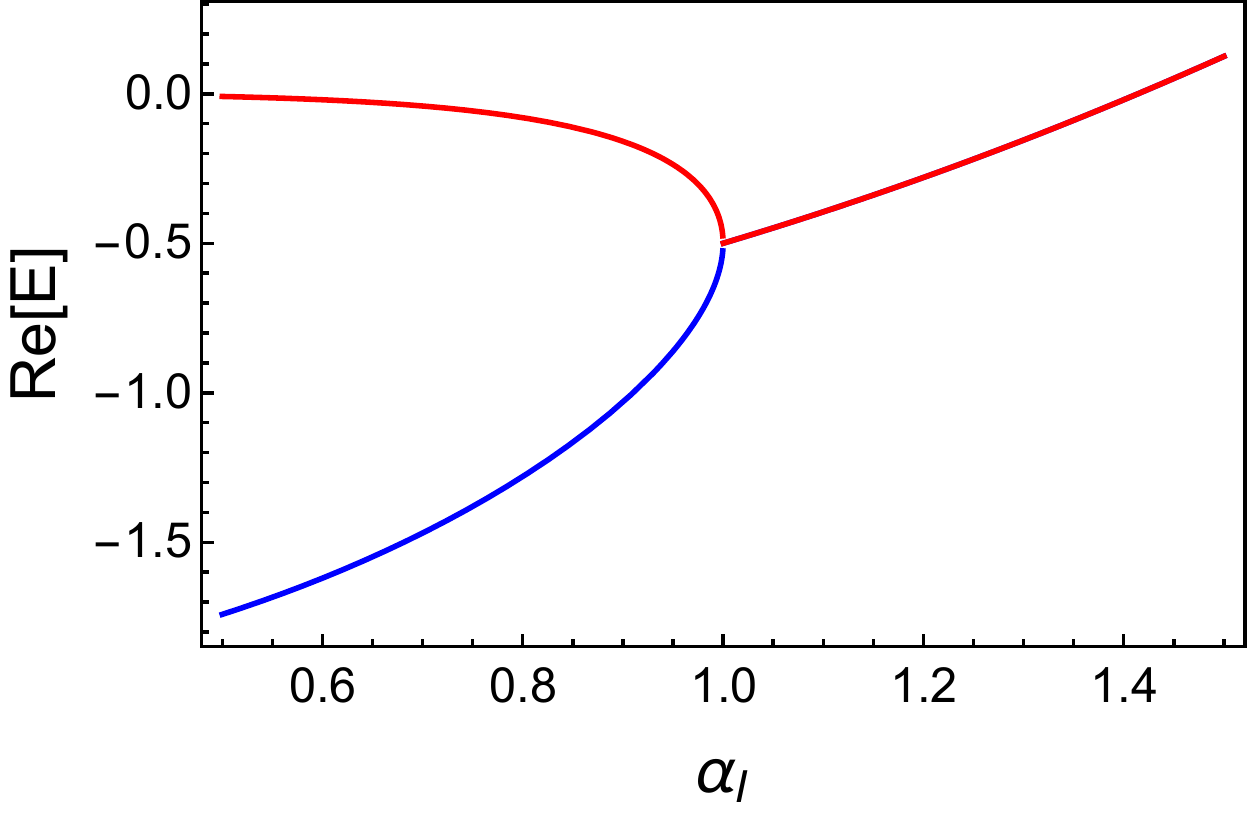}
\includegraphics[width=3.25in]{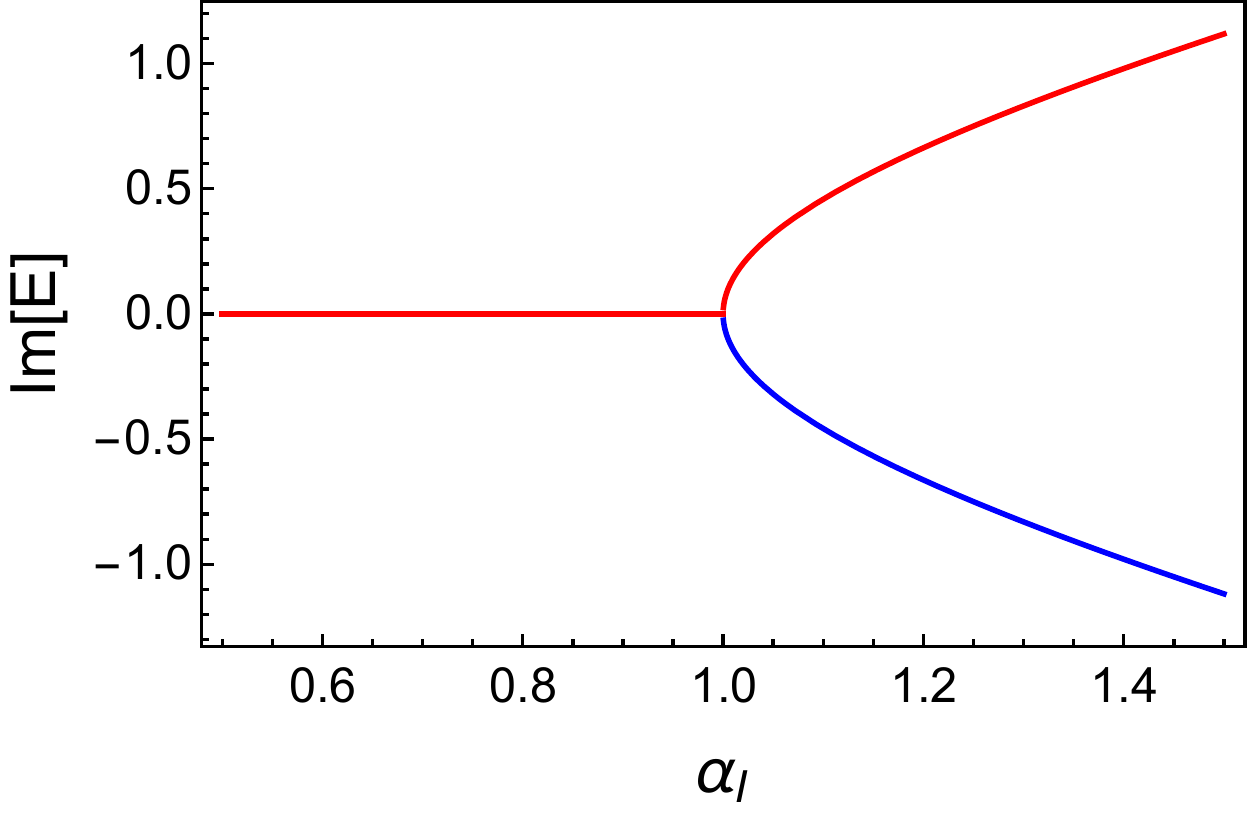}
\caption{${\cal PT}$ symmetry breaking for the ${\cal PT}$ symmetric delta function model. 
The two bound state eigenvalues are calculated using eq (\ref{eq:kappafloor}) with
$\alpha_R = -1$ and $\beta = 1$.
The upper plot shows the real parts of the two eigenvalues
and the lower plot shows the imaginary parts. The eigenvalues are plotted as a function of the
parameter $\alpha_I$. The eigenvalues are found to undergo a characteristic complementary pitchfork bifurcation.
Below the critical value $\alpha_I = 1$ the eigenvalues are real and above they are a complex conjugate
pair. The real parts of the eigenvalues degenerate at the critical value forming one pitchfork while the 
imaginary parts become non-zero forming the second complementary pitchfork.}
\label{fig:pitch} 
\end{figure}

In summary  $| \alpha_I | > 1$ and $ - \alpha_R /\beta > 0$ corresponds to the case of spontaneously 
broken ${\cal PT}$ symmetry. 
In this phase there are two bound states with complex conjugate energies. Otherwise ${\cal PT}$ 
symmetry is unbroken and there can be zero, one or two bound states all with real energy.
The behavior of the bound state energies across the ${\cal PT}$ symmetry breaking transition 
exhibits a characteristic complementary pitchfork form 
illustrated in fig \ref{fig:pitch}.

\subsubsection{Scattering states}

We now turn to the scattering of waves by a ${\cal PT}$ symmetric contact interaction.
In ${\cal PT}$ quantum mechanics the $S$-matrix is no longer unitary and hence its eigenvalues
are not required to be unimodular. When the eigenvalues are nonetheless unimodular 
${\cal PT}$ symmetry is said to be intact; when they cease to unimodular ${\cal PT}$ is said to be 
spontaneously broken. 

In order to determine the eigenvalues of the $S$-matrix we consider scattering states of the form
\begin{eqnarray}
\psi & = & A e^{i k x} + \lambda B e^{- i k x} \hspace{2mm} {\rm for} \hspace{2mm} x < 0,
\nonumber \\
& = & \lambda A e^{i k x} + B e^{- i k x} \hspace{2mm} {\rm for} \hspace{2mm} x > 0.
\label{eq:eigenscatter}
\end{eqnarray}
Here the amplitudes of the incoming waves from the left and right, denoted $A$ and $B$ 
respectively, are amplified by the eigenvalue $\lambda$ in the corresponding outgoing waves. 
Making use of the boundary condition eq (\ref{eq:gendelta}) yields two conditions connecting
the ratio $A/B$ and $\lambda$. Imposing consistency between these expressions reveals that 
the eigenvalues of the $S$-matrix are the roots of the quadratic equation
\begin{equation}
\Delta \lambda^2 + 4 i k  \cos \theta \lambda - \Delta^\ast = 0
\label{eq:seigenvalues}
\end{equation}
where 
\begin{equation}
\Delta = \gamma - \beta k^2 - i k (\alpha + \delta).
\label{eq:delta}
\end{equation}
In obtaining eq (\ref{eq:seigenvalues}) 
we have assumed that $\beta, \gamma$ and $(\alpha + \delta)$ are real and that $\alpha \delta - \beta \gamma = 1$. 
Hence our analysis to this point applies both to the hermitian and the ${\cal PT}$ symmetric contact interaction models. 

It is evident from eq (\ref{eq:seigenvalues}) that if $\lambda$ is an eigenvalue of the $S$-matrix
then so is $1/\lambda^\ast$. It is also clear that the product of the magnitudes of the two eigenvalues
must be 1. This also follows more generally because the $S$ matrix satisfies $S^\dagger S = {\cal I}$ 
and $S^\ast S = {\cal I}$ for the hermitian and ${\cal PT}$ symmetric cases respectively. (Here ${\cal I}$
denotes the $2 \times 2$ identity matrix.)

By writing down the explicit solution to eq (\ref{eq:seigenvalues}) it can be seen that 
if $| \Delta  |  > 2 k | \cos \theta | $ the eigenvalues are unimodular. On the other hand if $| \Delta | < 2 k | \cos \theta |$
the $S$-matrix eigenvalues no longer lie on the unit circle in the complex plane. One
has a magnitude bigger than unity; the other, smaller, in order to ensure that the product 
of the magnitudes is still unity. Physically one eigenmode of the $S$-matrix is amplified
upon scattering from the contact interaction; the other is attenuated. 

Making use of eq (\ref{eq:delta}) and exploiting $\alpha \delta - \beta \gamma = 1$ yields the useful formula
\begin{equation}
| \Delta |^2 - 4 k^2 \cos^2 \theta = (\gamma + \beta k^2)^2 + ( \alpha - \delta )^2 k^2 + 4 k^2 \sin^2 \theta. 
\label{eq:useful}
\end{equation}
From eq (\ref{eq:useful}) it is evident that in the hermitian case $\alpha = \delta$ and the right hand side is 
positive; hence $| \Delta | > 2 k | \cos \theta |$ always. In other words in the hermitian case we see by
explicit calculation that the eigenvalues of the $S$ matrix must be unimodular as expected
on general grounds also. However for the ${\cal PT}$ symmetric case $(\alpha - \delta)^2 = - 4 \alpha_I^2$
and hence the middle term on the right hand side of eq (\ref{eq:useful}) 
is negative. If it is sufficiently negative the eigenvalues of the
$S$ matrix no longer have to lie on the unit circle and ${\cal PT}$ symmetry is said to be broken. 
Eq (\ref{eq:useful}) reveals that there will always be a range of $k$ for which 
${\cal PT}$ symmetry is broken so long as $\alpha_I^2 > \sin^2 \theta$ and 
$\alpha_R^2 < \cos^2 \theta$. 

It is also interesting to examine the scattering amplitudes for the ${\cal PT}$ symmetric
contact interaction; for simplicity we only consider the case $b \neq 0$ and $\theta = 0$. 
If we consider a wave incoming from the left as in eq (\ref{eq:scatt}),
then the scattering amplitude $t$ is still given by eq (\ref{eq:tleftb}) but with $\kappa_\pm$
now given by eq (\ref{eq:kappafloor}). As for the hermitian case we see that the scattering
amplitude has poles in the upper half $k$-plane that are determined by the bound states 
if any. Furthermore in case the bound state energies are a complex conjugate pair the
scattering amplitude has a Lorentzian resonance whose location is determined by ${\rm Im} \; \kappa_\pm$
and width by ${\rm Re} \; \kappa_\pm$. This resonance has no counterpart in conventional
hermitian quantum mechanics. The effect of a non-zero $\theta$ on the scattering amplitude is
relatively innocuous; the amplitude is multiplied by $\exp ( \pm i \theta )$ depending on whether the
incident wave comes from $x \rightarrow \mp \infty$. 

\section{Kr\"{o}nig-Penney Model}

The Kr\"{o}nig-Penney model in one dimension is the simplest model of a crystal, originally introduced 
to provide a simple illustration of energy bands and band gaps in the early days of solid state physics. 
The model describes a particle that interacts with a periodic comb of delta functions
that are separated by a distance $\ell$. 
Here we consider two generalizations of the textbook model wherein the ordinary delta function is replaced
by either the generalized hermitian or the ${\cal PT}$ symmetric contact interactions introduced here. In the textbook case
an isolated attractive delta function would have a single bound state.
For a well separated array of delta function potentials this bound state fans out into a narrow band
characterized by the energy dispersion $E (k)$ where $E$ is the energy of the Bloch state 
and $k$ its crystal momentum which lies in the Brillouin zone $ - \pi/\ell < k < \pi/\ell$.

In the models considered here there can be two bound states in the isolated limit that fan into a 
pair of narrow bands when the contact potentials are well separated. We find that in the hermitian
case the gap between the bands can close when the parameters are tuned suitably. The energy 
dispersion near the intersection of the two bands is approximately that of a massless Dirac particle. 
In the ${\cal PT}$ symmetric case we find that
as the parameters are tuned the bands undergo a ${\cal PT}$ symmetry breaking transition. More precisely
recall that for an isolated contact interaction 
the ${\cal PT}$ transition takes place for $| \alpha_I | = 1$. For the corresponding Kronig-Penney
model we find that for $| \alpha_I | < \alpha_{c1} < 1$ the band energies $E_\pm (k)$ are entirely real; 
for $1 < \alpha_{c2} < | \alpha_I | $ the band energies become complex conjugate;
and for an intermediate range $ \alpha_{c1} < | \alpha_I | < \alpha_2$ the band energies are real for a small range of 
$k$ and complex conjugate elsewhere in the Brillouin zone. 

\subsection{Bloch analysis}

For a periodic potential in one dimension with a period $\ell$ the eigenfunctions must have the 
Bloch form
\begin{equation}
\psi_{nk} (x) = \Pi_{nk} (x) \exp ( i k x )
\label{eq:bloch}
\end{equation}
Here $n$ is an index that labels the bands and $k$ is the crystal momentum which lies in the 
Brillouin zone $ - \pi/\ell < k < \pi/\ell$. The form factor $\Pi_{nk}$ is a periodic function of $x$
with period $\ell$. Hence the Bloch wave-function obeys the quasi-periodic condition
\begin{equation}
\psi_{nk} (x + \ell) = \psi_{nk} (x) \exp( i k \ell )
\label{eq:quasiperiod}
\end{equation}
In the Kr\"{o}nig-Penney model considered here we assume that the particle experiences
a contact interaction at the points $x = \nu \ell$ where $\nu$ is an integer. This includes the origin 
which corresponds to $\nu=0$. 
Hence $\psi_{nk}$ must obey the boundary condition eq (\ref{eq:gendelta}) at the origin.
Since we are interested in bound state bands with negative energy we take the 
Bloch wave-function to have the form
\begin{eqnarray}
\psi_{nk} & = & A \exp ( \kappa x ) + B \exp ( - \kappa x ) \hspace{3mm} {\rm for} \hspace{2mm} - \frac{\ell}{2} < x < 0,
\nonumber \\
& = & C \exp ( \kappa x ) + D \exp ( - \kappa x ) \hspace{3mm} {\rm for} \hspace{2mm} 0 < x < \frac{\ell}{2}.
\nonumber \\
\label{eq:blochansatz}
\end{eqnarray}
The energy of this state is $ E = - \hbar^2 \kappa^2 / 2 $. 

Imposing the quasi-periodicity condition (\ref{eq:quasiperiod}) and the boundary condition (\ref{eq:gendelta}) 
leads to the quantization condition
\begin{eqnarray}
\kappa^2 \beta + \kappa (\alpha + \delta) + \gamma & = & 
4 \kappa \cos ( k \ell - \theta ) e^{- \kappa \ell} 
\nonumber \\
& + & 
[ \kappa^2 \beta - \kappa ( \alpha + \delta ) + \gamma ] e^{-2 \kappa \ell }.
\nonumber \\
\label{eq:quantization}
\end{eqnarray}
Our task now is to solve the transcendental equation
(\ref{eq:quantization}) for $\kappa$. By determining the dependence
of $\kappa$ on $k$ we can determine the energy dispersion $E(k)$. 
In the limit $\ell \rightarrow \infty$ the right hand side of eq (\ref{eq:quantization}) vanishes and the
allowed $\kappa$ values are the same as for an isolated contact interaction as expected. 
The analysis for finite $\ell$ is undertaken separately below for the hermitian and ${\cal PT}$ symmetric
cases. In both cases for simplicity we will take $\theta = 0$ since the only effect of non-zero $\theta$ is
to shift the bands in $k$-space. 

\subsection{${\cal PT}$ symmetric bands}

Recall that for the isolated ${\cal PT}$ symmetric contact interaction the allowed $\kappa$ values are
given by eq (\ref{eq:kappafloor}). Since we are interested in the ${\cal PT}$ symmetry breaking
transition we consider values of $| \alpha_I |$ near to the threshold value of unity. For brevity we write $
\kappa_\pm = \overline{\kappa} \pm i \epsilon$ for $| \alpha_I | > 1$ and $\kappa_\pm = \overline{\kappa} 
\pm \epsilon$ for $ | \alpha_I | < 1$ respectively. We assume that $\ell$ is sufficiently large that the bands
will be narrow and hence posit that $\kappa = \overline{\kappa} + \Delta$ where $\Delta$ is small. To enforce
that the bands are narrow we assume $\exp( - \kappa \ell ) \ll 1$ and we also assume that $\Delta \ell \ll 1$. 
The condition $\exp ( - \kappa \ell ) \ll 1$ allows us to neglect the second term on the right hand side of
the quantization condition and the condition that $\Delta \ell  \ll 1$ allows us to approximate 
$\exp( - \kappa \ell ) \approx \exp ( - \overline{\kappa} \ell )$. Making these approximations
yields
\begin{equation}
\Delta_\pm (k) = \pm W \sqrt{ \varepsilon^2 + \cos k \ell }
\label{eq:belowtrans}
\end{equation}
for $| \alpha_I | < 1$ and 
\begin{equation}
\Delta_\pm (k) = \pm W \sqrt{ \cos k \ell - \varepsilon^2 }
\label{eq:abovetrans}
\end{equation}
for $| \alpha_I | > 1$. 
Here for simplicity we have defined 
\begin{equation}
W =  \left[ \frac{4 \overline{\kappa} }{\beta} \exp ( - \overline{\kappa} \ell ) \right]^{1/2} 
\label{eq:width}
\end{equation}
and $\varepsilon$ is a rescaled version of $\epsilon$ given by 
$\epsilon = W \varepsilon$. 

\begin{figure*}
\centering
\begin{tabular}{cc}
\includegraphics[width=70mm]{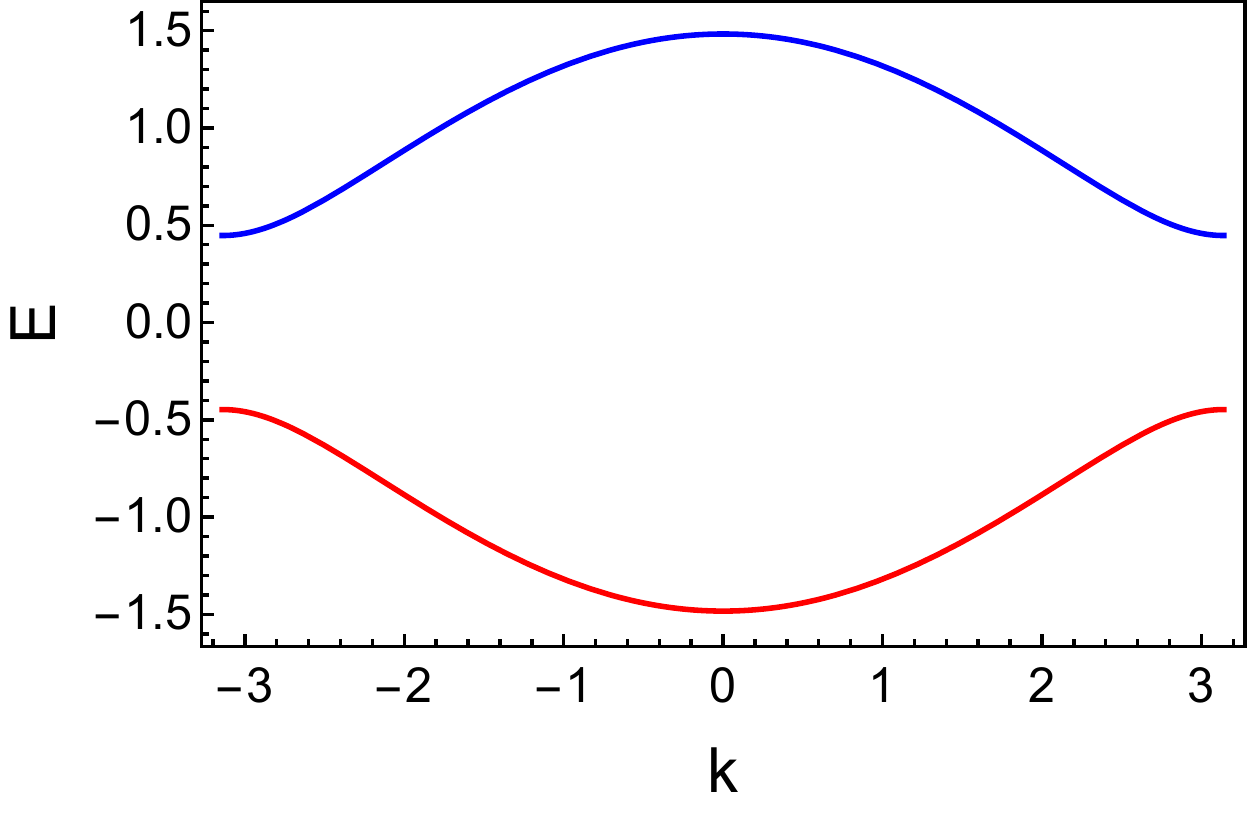}&
\includegraphics[width=70mm]{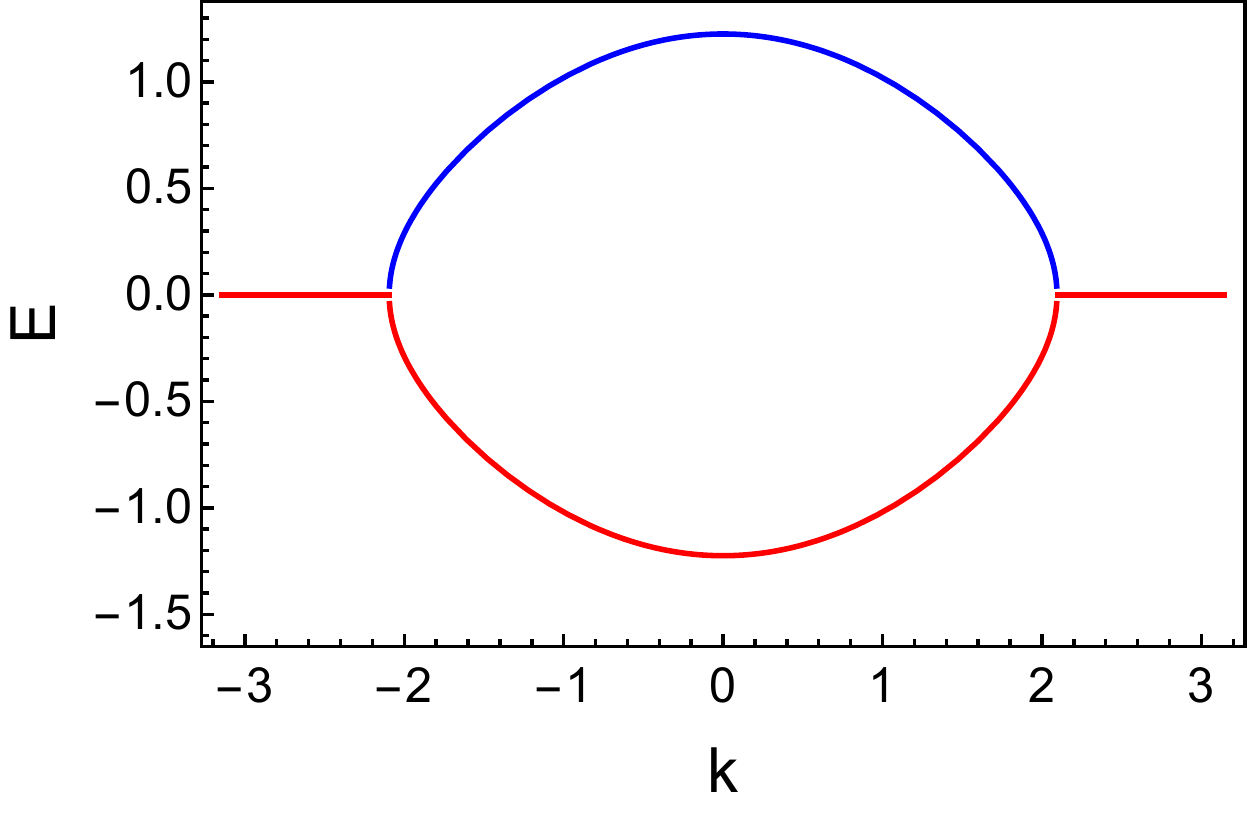}\\
\includegraphics[width=70mm]{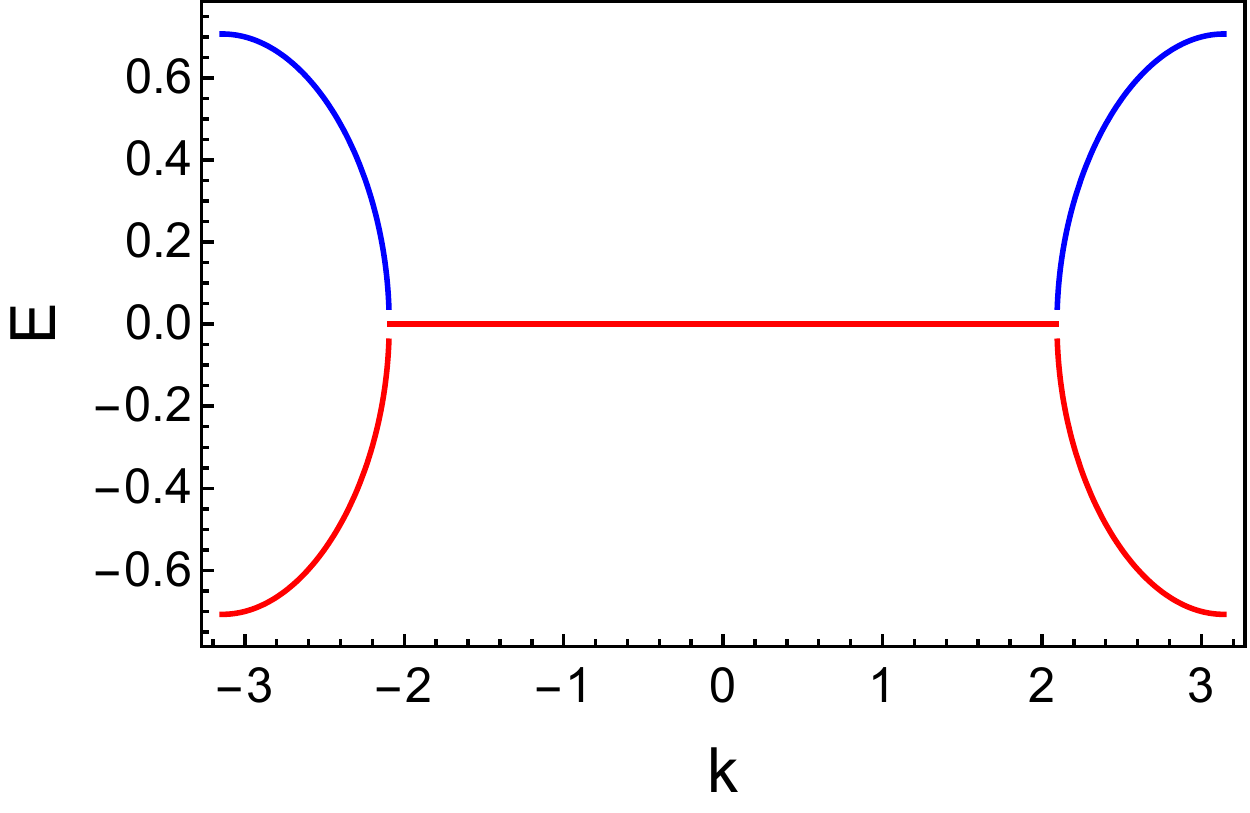}&
\includegraphics[width=70mm]{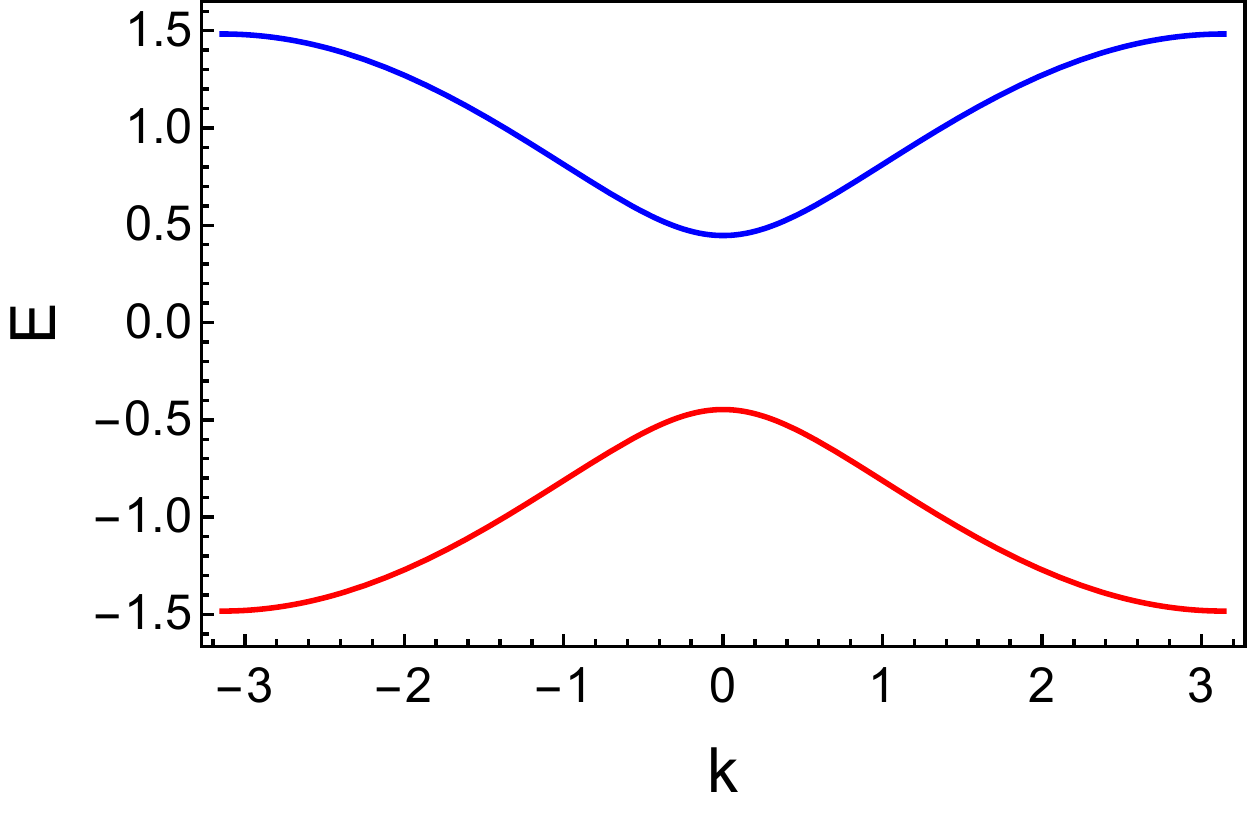}\\
\end{tabular}
\caption{The energy bands of the ${\cal PT}$ symmetric Kr\"{o}nig-Penney model. The plot on top left shows
the energy plotted as a function of wave-vector for the case of unbroken ${\cal PT}$ symmetry.
This is the case (i) in the main text. The energy bands in this case are entirely real and
have an approximate particle-hole symmetry. 
The plot on bottom right by contrast corresponds to case (iv) where the ${\cal PT}$ symmetry breaking
transition is complete. The energy bands are a complex conjugate pair. The plot only shows the dispersion
of the imaginary parts of the energy because the real part of the energies is found to be constant. 
The symmetry between the bands in this case is required by ${\cal PT}$ symmetry. The plots on 
top right and bottom left correspond to the onset of the ${\cal PT}$ transition. This is case (iii) in the
text. The plot on top right shows the real part of the energy and the plot on the lower right shows the
imaginary part. The bands have been offset so as to be centered about zero energy and the parameter
$W = 1$ and $\ell = 1$ throughout whereas $\varepsilon^2 = 1.2$ (top left and bottom right) or $\varepsilon^2 = 0.5$ (top
right and bottom left).}
\label{fig:bandbreak}
\end{figure*}

In summary the energy bands near the ${\cal PT}$ transition in the narrow band limit are
given by the simple expression
\begin{equation}
E_\pm (k) = - \frac{1}{2} \overline{\kappa}^2 - \overline{\kappa} \Delta_\pm (k)
\label{eq:ptbands}
\end{equation}
where $\Delta_\pm (k)$ is given by eq (\ref{eq:belowtrans}) and (\ref{eq:abovetrans}) for the cases
$| \alpha_I |< 1$ and $| \alpha_I | > 1$ respectively. $W$ is a measure of the bandwidth and $\varepsilon$
measures the distance of $|\alpha_I |$ from the transition value of unity. It is evident from these expressions
that there are four regimes. (i) For the ${\cal PT}$ symmetric regime $| \alpha_I | < 1$ and $\varepsilon > 1$ and
the bands are pure real. (ii) For the broken ${\cal PT}$ symmetry regime $| \alpha_I | > 1$ and $\varepsilon > 1$
the two bands are a complex conjugate pair. (iii) The range $| \alpha_I | < 1$ and $\varepsilon < 1$ corresponds
to the onset of the ${\cal PT}$ transition. In this regime the bands are real for small $k$ and complex conjugate
elsewhere in the Brillouin zone. (iv) The range $| \alpha_I | > 1$ and $\varepsilon < 1$ corresponds to the 
range over which the ${\cal PT}$ transition is completed. Over this range too the bands are partially real
at small $k$ and complex conjugate elsewhere in the Brillouin zone. These behaviors are shown in 
fig \ref{fig:bandbreak}.  

\subsection{Hermitian bands}

We now analyze the quantization condition eq (\ref{eq:quantization}) for the hermitian
case. 
We focus on the case that for an isolated contact interaction there
are two bound states. It is more convenient to work with the parameters $\kappa_1, \kappa_2$ and $\beta$
that were introduced at the end of section \ref{sec:model}
to describe the contact interaction. In terms of these
parameters the exact quantization condition (\ref{eq:quantization}) 
may be rewritten
\begin{equation}
(\kappa - \kappa_1) (\kappa - \kappa_2) = \frac{4 \kappa}{ \beta } \cos k \ell e^{-\kappa \ell} +
(\kappa + \kappa_1)(\kappa + \kappa_2) e^{-2 \kappa \ell}.
\label{eq:secondquantization}
\end{equation}

First for simplicity we assume that the two bound states are well separated in comparison
to the width of the bands that they form. Now in order to analyze the band associated with
the first isolated bound state we write $\kappa = \kappa_1 + \Delta$ where $\Delta$ is
assumed to be small. More precisely we assume that $\Delta \ell \ll 1$ and also that
$\exp ( - \kappa_1 \ell ) \ll 1$. The latter assumption allows us to ignore the second
term on the right hand side of eq (\ref{eq:secondquantization}) and we obtain 
\begin{equation}
\Delta = \frac{ 4 \kappa_1 }{\beta ( \kappa_1 - \kappa_2 ) } e^{ - \kappa_1 \ell } \cos k \ell.
\label{eq:upperdelta}
\end{equation}
Recalling that the energy is given by $E = - \kappa^2 / 2$ we find that the first
band has the energy dispersion 
\begin{equation}
E_1 (k) = - \frac{ \kappa_1^2 }{2 } - \frac{4  \kappa_1^2}{\beta ( \kappa_1 - \kappa_2 ) } 
e^{- \kappa_1 \ell} \cos k \ell
\label{eq:upperband}
\end{equation}
Similarly the second band is given by
\begin{equation}
E_2 (k) = - \frac{ \kappa_2^2 }{2} + \frac{4 \kappa_2^2}{\beta ( \kappa_1 - \kappa_2 ) } 
e^{- \kappa_2 \ell} \cos k \ell
\label{eq:lowerband}
\end{equation}

More interesting behavior results when $\kappa_1 \approx \kappa_2$. In this regime as the parameters
of the generalized delta function potential are tuned appropriately the bands intersect at an isolated 
point in $k$-space before moving apart again. At the intersection the bands form a Dirac cone. 
To demonstrate this behavior we write $\kappa_1 = \overline{\kappa} + \epsilon$ and $\kappa_2 =
\overline{\kappa} - \epsilon$ where $\epsilon$ is positive and assumed to be small in a sense
to be made precise. As before we 
write $\kappa = \overline{\kappa} + \Delta$ where $\Delta$ is also assumed to be small. For simplicity
we assume that the delta potentials are well separated, $\exp ( - \overline{\kappa} \ell ) \ll 1$, 
but that $\Delta$ is sufficiently small that $\Delta \ell \ll 1$. Due to the constraint
$( \kappa_1 - \kappa_2) | \beta | = 2 \epsilon | \beta | \geq 2$, a small value of $\epsilon$ implies a
large value of $\beta$; hence we can no longer neglect the second term on the right hand side
of eq (\ref{eq:secondquantization}) in comparison to the first. In fact the two terms are of the same
order if we take 
\begin{equation}
\epsilon = 2 \overline{\kappa} e^{- \overline{\kappa} \ell } \varepsilon
\label{eq:varepsilon}
\end{equation}
where $\varepsilon$ is of order unity. We also write $1/\beta = f \epsilon$ with $-1 < f < 1$ (in order to
respect the constraint $| \beta | \epsilon > 1$). 
Making these assumptions and using eq (\ref{eq:secondquantization}) 
we obtain
\begin{equation}
\Delta = \pm 2 \overline{\kappa} e^{ -\overline{\kappa} \ell } \left( 1 + \varepsilon^2 + 2 \varepsilon f \cos 
k \ell \right)^{1/2}.
\label{eq:deltaclosure}
\end{equation}
This corresponds to the energy bands
\begin{equation}
E_{\pm} (k) = - \frac{\overline{\kappa}^2}{2} \mp 2 \overline{\kappa}^2 e^{ - \overline{\kappa} \ell } 
\left[ 1 + \varepsilon^2 + 2 \varepsilon f \cos k \ell \right]^{1/2}.
\label{eq:bandgap}
\end{equation}

From eq (\ref{eq:bandgap}) we see that for
$\varepsilon = 1$ and $f = -1$ or $f = +1$ the bands touch at $k = 0$ or $k = \pi$ respectively. 
For $k$ near the intersection the energy dispersion is approximately linear and the bands
form a massless Dirac cone. In terms of the original parameters $\varepsilon = 1$ and $f = \pm 1$
translates to $( \kappa_1 - \kappa_2 ) | \beta | = 2$ where the sign of $\beta$ is the same as that
of $f$. Put another way there are two gapped phases corresponding to $ \beta > 2/(\kappa_1 - \kappa_2)$
and $\beta < - 2/(\kappa_1 - \kappa_2)$ respectively. 

\section{Conclusion}

The delta function potential is a simple model of zero range contact interaction in one dimension. In this paper we have
introduced generalizations of the delta function for conventional hermitian quantum mechanics and
${\cal PT}$ quantum mechanics in one dimension. We find that the corresponding generalizations of the Kr\"{o}nig-Penney
model exhibit interesting behavior in both hermitian and ${\cal PT}$ quantum mechanics. In ${\cal PT}$
quantum mechanics we find bands that undergo ${\cal PT}$ symmetry breaking, providing a particularly
simple example of this phenomenon. 
In hermitian quantum mechanics we find that the gap between the two bound state bands closes
when the parameters of the interaction are appropriately tuned yielding a conical intersection between
the bands at a single point in the Brillouin zone. Near the intersection the dispersion relation is that
of a massless Dirac fermion. Whether the gapped phase on either side of gap closure is a topological
insulator is an intriguing question we leave open for future work. 

Another interesting application of our generalized contact interaction may be to many body physics in
one dimension. There are only a handful of exactly soluble non-trivial models of quantum many body systems. 
In a seminal paper Lieb and Liniger showed that a one dimensional gas of bosons interacting 
via a delta function contact interaction 
was soluble via Bethe ansatz \cite{lieb, mattis}. 
There has been a resurgence of interest in this class of integrable models due to their experimental 
realization in cold atoms
\cite{coldscience, coldnature, coldatoms}
A natural generalization of the Lieb-Liniger model suggested
by this paper is to replace the delta function interaction with the generalized form of contact interaction
studied here. This model also should be soluble via Bethe ansatz and may be realizable with cold atoms. 

\vspace{3mm}

{\em Acknowledgement.}
Kristin McKee was supported by SURES, a summer undergraduate research program of Case
Western Reserve University.


\begin{thebibliography}{999}
\bibitem{hermiticity} In this context we use ``hermitian'' synonymously with the more precise term
``self-adjoint''; see section \ref{sec:hermitdelta}.
\bibitem{footnote:gauge} Naively one might suppose that it is possible to eliminate the phase $\theta$ by making 
the following gauge transformation: $\psi \rightarrow e^{- i \theta} \psi$ for $ x > 0 $ and $\psi \rightarrow \psi$ for 
$ x < 0$. However even for an open one dimensional system physics is invariant only under small gauge
transformations $\psi \rightarrow \psi e^{ i \gamma}$ where $\gamma \rightarrow 0$ for $x \rightarrow \pm \infty$. 
Indeed we will see below that although the bound states are independent of $\theta$ the scattering matrix
does depend on $\theta$ in a non-trivial way. 
This is a particularly simple example of the subtle distinction between large and small
gauge transformations in quantum mechanics.
\end{thebibliography}
\end{document}